# Born to be Wild: Using Communities of Practice as a Tool for Knowledge Management

By Valérie Chanal and Chris Kimble

**Abstract**
This paper looks at what happens when Communities of Practice are used as a tool for Knowledge Management. The original concept of a Community of Practice appears to have very little in common with the knowledge sharing communities found in Knowledge Management, which are based on a revised view of 'cultivated' communities. We examine the risks and benefits of cultivating Communities of Practice rather than leaving them 'in the wild'. The paper presents the findings from two years of research in a small microelectronics firm to provide some insights into the wild vs domesticated dichotomy and discusses the implications of attempting to tame Communities of Practice in this way.

## 1   Introduction

This paper is concerned with what happens when groups known as Communities of Practice are used as a tool for Knowledge Management. Recently there has been a noticeable move toward the development of what are termed knowledge sharing communities, which has been linked by some to the failure of more traditional IT based approaches to sharing or distributing knowledge. Communities of Practice, it is argued, can provide the solution to this problem.

The term Communities of Practice was first coined as part of a theory of situated learning by Jean Lave and Etienne Wenger almost 20 years ago in their book "*Situated Learning: Legitimate Peripheral Participation*" (Lave & Wenger, 1991). Since then, Communities of Practice have also become the focus of attention in the field of Knowledge Management. Seen from the Knowledge Management perspective, ICT tools deal with the more easily captured explicit knowledge, while Communities of Practice provide the solution to the management of the more problematical tacit knowledge, which, because it cannot be transferred directly, is a source of competitive advantage (Nonaka, 1994).

In his book "*Cognition In The Wild*", Hutchins (1995) uses the term 'wild' to refer to human cognition in its natural setting; that is, a situated and socially constituted activity as opposed to the artificial setting of the laboratory. Drawing a similar distinction, we describe a shift from the Communities of Practice described in the early works - communities that are essentially 'in the wild' - to the cultivated and controlled groups in the later works - in effect, communities that have been domesticated. In this paper, we ask if the risks and benefits of cultivating Communities of Practice bring more benefits than leaving them 'in the wild'.

To answer this question, we will first review some of the literature on Communities of Practice to show how the concept has changed and to delineate the distinction between wild and domesticated communities. We then continue with a review of the literature relating to Communities of Practice and Knowledge Management. We end the literature review with a summary of the challenges that are faced when Communities of Practice are created as part of a Knowledge Management initiative within a host organization.

The empirical section of the paper will use the findings from research carried out in a small microelectronics firm in France, e2V, over a period of two years (Cappe, 2008). We present the results from two experimental Communities of Practice that were created within the company, and in doing so, provide some additional insights into the wild vs domesticated dichotomy. We conclude by examining the risks and benefits of treating Communities of Practice in this way; a discussion of the implications of attempting to tame Communities of Practice and an indication of how future work in this area might be developed.

## 2    Communities of Practice - The Evolution of an Idea

The term Community of Practice has been used in a variety of ways. Adopting the approach of Cox (2005) we will contrast the approach in the early works, where Communities of Practice are seen as emergent and creative groups, to that of later works, where they are seen as groups that can be cultivated to serve the needs of a host organization.

### 2.1    Communities of Practice as Emergent and Creative Groups

In "*Situated Learning: Legitimate Peripheral Participation*" (Lave & Wenger, 1991), Lave and Wenger outlined an alternative to the behaviourist theories of learning that were dominant at the time. Their description of Communities of Practice provides an account of how situated learning takes place through being enacted in practice. Learning and practice mutually shape one another in a continuous and iterative social process, i.e. they are "*mutually constitutive*" (Lave & Wenger, 1991, p. 177) . The communities they describe have no fixed structure and change gradually over time with the ebb and flow of changing membership. Lave and Wenger's view of learning focused on its socially negotiated nature. Learning is seen as part of the process of socialization into a community: a newcomer only becomes a full member of the community through gradually learning its practice, language and conventions. Membership creates a sense of identity both in the eyes of the outside world (through being associated with the community) and within the community (through the degree to which one's skill and knowledge is recognized by others in the community).

Lave and Wenger (1991) use the concept of Legitimate Peripheral Participation (LPP) to explain how members move between the core and the periphery. Legitimation and participation together define different ways of belonging to a community, whereas peripherality and participation are concerned with identity in the social world. Brown and Duguid (1991) illustrate this by using the stories told by the Xerox tech-reps to illustrate how, through the telling and re-telling of these stories, the tech-reps become a "*community of interpretation*" (Brown & Duguid, 1991, p. 47). When viewed in this way, the development of the community can be viewed as an ongoing performance: an improvisation that is enacted and re-enacted by the members of the community as they go about their daily activities. Thus the  learning that takes place in Communities of Practice is not just situated learning but *"generative social practice"* (Lave & Wenger, 1991, p. 35) that can change lives.

For example, based on observations of Alcoholics Anonymous meetings, Lave and Wenger describe (1991, pp. 79 - 84) how the practice of an Alcoholics Anonymous meeting is effectively the creation of an identity of a 'Non Drinking Alcoholic'. In Alcoholics Anonymous meetings, members stand up and tell stories of their past lives for others in the meeting. These stories act a model of the behaviour of an alcoholic. The hope being that members who have yet to come to terms with their own alcoholism will find so much of their lives in these stories, that they will ask if they too are alcoholics. Thus, for newcomers, the members' stories are not simply a description of the life of an alcoholic, but provide a means to reinterpret their past and create a future in terms of their new identity of an alcoholic.

## 2.2   Communities of Practice as Cultivated and Constrained Groups

The more recent works on Communities of Practice, such as "*Cultivating Communities of Practice*" (Wenger, McDermott, & Snyder, 2002), are aimed primarily at practitioners; here the emphasis is on ways to manage the community and the role it can play within an organization.  Wenger, McDermott and Snyder state explicitly "*... we have concentrated primarily on the ability of Communities of Practice to steward knowledge inside organizations*" (Wenger et al., 2002, p. 219) as "*... they do not merely manage knowledge assets: they create value in multiple and complex ways*" (Wenger et al., 2002, p. 215).

In these later works, Wenger abandons the notion of social communities based on LPP and adopts a different view of Communities of Practice.  This new vision is based on the notion of sense-making in organizations and the concept of dualities, which he describes as, *"... a single conceptual unit that is formed by two inseparable and mutually constitutive elements*" (Wenger, 1998a, p. 66).  He describes the forces that motivate the community in terms of the tensions that exist within and between dualities, and identifies four such dualities: participation-reification, designed-emergent, identification-negotiability and local-global.  Of the four, the participation-reification duality, with its close links to Knowledge Management, has been the focus of the greatest interest.

While some link the participation-reification duality to notions such as Nonaka's tacit and explicit knowledge, such comparisons can be misleading.  For Nonaka, tacit and explicit knowledge are seen as distinct forms of knowledge, although one may be 'converted' to another through a cycle of socialization, externalization, combination and internalization known as the SECI model (Nonaka, 1994).  For Wenger however, the tacit/explicit dichotomy is a false dichotomy, because all knowledge is formed simultaneously through both participation and reification: each is reflected in the other.

Finally, these new cultivated communities have lost the autonomy and freedom that was present in the earlier descriptions of communities 'in the wild'.  This later view of Communities of Practice sees them as groups that can be intentionally cultivated by providing appropriate managerial inputs.  For example, Snyder and Briggs note that while Communities of Practice are still essentially informal structures "*sponsors and stakeholders have important roles to play*" (Snyder & Briggs, 2003, p. 7).  Communities of Practice are now simply "*a different cut on the organization's structure*" (Wenger, 1998a) that arise out of a need to accomplish a particular task in the organization.

## 2.3   Communities of Practice and Knowledge Management: The Challenges

As we noted previously, Communities of Practice have long been the focus of interest among sections of the Knowledge Management community.  Developments in Information Technology, coupled with awareness of the importance of organizational knowledge, have led to the development of a variety of Information Systems to manage knowledge.  However, while IT has proved successful at managing some types of knowledge, the hard to capture tacit knowledge that Nonaka (1994) argues is the basis for competitive advantage, remains anchored in individuals.  Much of the literature dealing with this area is written from the viewpoint that Communities of Practice can provide a suitable environment to share or exchange knowledge between different groups in an organization (Zboralski, 2009).

Although it is possible to make conceptual links between Communities of Practice and the management of tacit knowledge, the Communities described by Lave and Wenger (1991) seem ill-suited to the task. Wenger, McDermott and Snyder (2002) provide an alternative view of Communities of Practice that is more amenable to this viewpoint, but it is not without its problems. Wenger notes that, "*Communities of Practice give you not only the golden eggs but also the goose that lays them* [but] *the challenge for organizations is to appreciate the goose and to understand how to keep it alive and productive*" (Wenger & Snyder, 2000, p. 143). Similarly, Brown and Duguid note that attempts to control or organize Communities of Practice will only succeed in disrupting them (Brown & Duguid, 1991, p. 49). Examples of this can be found in empirical studies such as Gongla and Rizzuto (2004) who note that if an organization 'spotlights' a Community of Practice, *"... the community may remove itself completely from the organizational radar... pretending to disperse, but in reality continuing to function outside of the organization's purview*" (Gongla & Rizzuto, 2004, p. 299).

In the next section, we will present results from two experimental Communities of Practice that were intentionally cultivated in a company. In particular, we will look at the extent to which organizations are able to instrumentalize Communities of Practice and at the risks they run when attempting to do so. We ask, in terms of desired strategic outcome of managing tacit knowledge, does the instrumentalization of Communities of Practice risk killing the goose that lays the golden eggs?

## 3 Case study: A Process of Cultivating Two Communities of Practice

The data for this study was collected between January 2005 and December 2007 in e2V Grenoble, a subsidiary of the e2V Group, consisting of 480 employees, of whom 250 were highly qualified engineers and managers. The main activity of this subsidiary is the design and testing of microelectronics systems for the medical, telecommunications, automotive and aerospace markets. The company is organized into separate business units, each of which has their own marketing, design, engineering and quality control activities. The methodological approach can be characterized as "*recherche ingénierique*" (Chanal, Lesca, & Martinet, 1997). This approach is similar to action research, in that it is concerned with the researcher's active involvement in the processes of organizational change; however, it is distinguished by the creation of a "*chercheur-ingénieur*" (researcher-engineer) who designs a tool to support their research, builds it and acts as a moderator and evaluator of its implementation.

### 3.1 The Failure of the Traditional Approaches to Knowledge Management

An initial field study, in the form of a diagnostic examination of the existing systems used for Knowledge Management, was undertaken in 2005. The study focused on the four main tools used by the management of the company to manage and retain knowledge.

1. An intranet system for the distribution of technical knowledge
2. A document management system for tracking issues related to quality
3. A 'dual ladder' system of promotion for technical experts who do not normally take management responsibilities
4. A phase in the quality management process termed "*retour d'expérience*" (REXP)

The results of the study were not encouraging for a company that believed it was actively managing its knowledge.

The intranet aimed to provide basic information throughout the company, such as the technical description of products, learning guides, directories and so on.  However, this was not widely used as people found it difficult to apply the abstracted, canonical knowledge it contained to other contexts.  The document management system was not much used outside the group of people who produced the documentation.  The technical experts, although having a specific and unique role within the organization, where not often consulted by those outside of their normal working environment.  Finally, project leaders were required to fill in a retour d'expérience (REXP) form that was supposed to help others to capitalize on the experience that has been gained from each project.  However, only 20 % of the projects ever returned a correctly documented REXP form.  The results of the study highlighted the limits of traditional Knowledge Management tools to facilitate knowledge sharing; the management of the company rethought its position and began to look for solutions that were more practice-based.  By way of an experiment, they hired a student who had just completed her masters' degree with the company, and was now starting her PhD, to act as a Knowledge Manager.  It is her work (Cappe, 2008) that forms the bulk of the data presented in this paper.

Cappe carried out more than 70 interviews with engineers to identify specific practices and areas in need of knowledge sharing.  Her study revealed a general need to share knowledge across the organization.  Using a set of basic criteria to characterize a Community of Practice, she identified two categories of people, who would, in her view, benefit from this approach.  Two Communities of Practice were created: one for project leaders and one for scientific experts.  The previous study had helped to identify people who would be both highly motivated and be seen as having the legitimacy / informal authority to bring in others to who would participate.  The experiment also required the support of the management of the company, and consequently a steering committee was set up to oversee the experiment.

### 3.2   The Experimental Communities of Practice
Below, we will briefly provide some contextual / background information on the two communities, before presenting our results.

**The Project Leaders**
The experiment to create a Project Leaders Community of Practice involved 8 full time and 20 part time project leaders in the company.  All 28 shared a common area of work, although most of them belonged to different business units and had few opportunities to meet.

Three potential brokers were identified, and in a preliminary meeting before the official launching of the experiment, they all expressed their wish to improve the sharing of knowledge.  Following an initial meeting, where project leaders were given feedback on the diagnosis phase of the research, 20 people agreed to participate in the experiment.  Seven meetings were organized in the first year (2005) and ten in the second year (2006) with an average participation of around 15 members per meeting.

**The Scientific Experts**
Under the company's dual ladder promotion policy, it had assigned expert status to 18 people who had specific technical expertise that was crucial to the success of the company.  Despite the fact that each expert had a short presentation on the company's intranet, the majority of them knew nothing about the other experts in the organization.

As for the project leaders, a first meeting was organized with three potential brokers and it was suggested that the Vice President of Strategy and Business Development (a member of the executive committee) should act as a broker. During this first meeting, all three experts confirmed the need to improve mutual knowledge and knowledge sharing.

Fifteen experts decided to join the community, although some of them expressed doubts about the value of the experiment. They decided to start with the objectives of getting to know each other better and working towards the wider acknowledgment of expert status within the company. Four meetings were organized in the first year, but only one in the second year.

## 4 The Actions Taken by the Two Communities

In this section, we will present some examples of actions taken by the communities. We will show how these actions led to the communities renegotiating certain rules with the managers of the company. Our observations are organized around the following two themes:

1. The practice of the community, i.e. data was drawn from the minutes of the meetings and from direct observation of groups.
2. The evaluation of the experiment by the members themselves and by the executive committee, i.e. data was collected in interviews conducted at the end of the study.

### 4.1 The Project Leaders Community of Practice

During the first meetings, the members defined the objectives of the community: a benchmarking of project management methods, a sharing of experiences and proposals to improve practices. After the steering committee had approved these objectives, the members began to share experiences about how they led projects and how they coped with day-to-day difficulties. A common problem was that changes to technical requirements during a project frequently led to tensions between the project leader and the marketing department. The marketing department felt under pressure from the customer to agree to changes to a product's requirements without changes to the initial cost and without incurring any delay. However, any change in the requirements had an impact on the overall management of the project. Although the cost and delay of re-evaluations was allowed for, the tension between these two departments tended to lead to a drift away from the initial objectives of the project.

During these sessions, it appeared that there was a particular difficulty related to the production phase. The engineer in charge of production did not participate in any of the upstream phases and so was not able to indicate the constraints faced during production. The proposed solution was that there should be a new milestone called 'start of industrialization'. The community asked the Director in charge of quality to participate in a meeting to discuss this issue with them. After some discussion, the principle of adding this milestone was approved. The job description files of the product engineers were modified to take into account this new milestone. This had a positive effect on the motivation of the group, who were able to see the results of their actions expressed in the official project management processes of the company.

### 4.2 The Technical Experts Community of Practice

In their first meetings, the experts presented summaries of their activities and domain of expertise. The informal discussions had revealed problems concerning the definition of an expert's duties and the time that they could devote to these duties. An example of this was that, if experts asked for an account number to charge for the time that had been dedicated to a particular project, they would need to explain the precise nature of their duties and justify the amount of time that should be allocated to it.  For the experts, the need to go through this process each time they were consulted highlighted the need for some form of official recognition of the nature of their role in the company.  The community decided to focus on this issue and to try to improve the visibility of the expert's role within the organization.  A second example concerned a proposal that was put forward by the experts to create a library of standard technology modules.  Such a project would require both time and financial resources; however, the steering committee decided it was not a priority and refused to allocate the necessary resources.  The lack of support for this proposal led to a disengagement of the members of the community and, in year 2, only one meeting took place.

In contrast, a different episode of the life of the community showed how support and recognition from the organization could lead to improved motivation within the community.  Six months after the launching, the expert's community received a request from the executive committee to draw a map of key knowledge domains within the company in order to prepare a long-term strategic plan.  During this work, some important strategic points were highlighted, such as the possibility of merging certain manufacturing processes.  This contribution by the community led to the executive committee redefining and enlarging the formal role of the experts, which had previously been limited to technical problem solving.  This had a positive effect on the motivation and the cohesion of the members, and the experts who had not initially participated in the community, finally decided to join.

## 5   The Evaluation of the Two Communities of Practice

The previous section has illustrated some of the positive aspects of the two experiments of cultivating Communities of Practice.  They contributed to the sharing of knowledge and practices across the traditional organizational boundaries, however they also generated tensions.  These tensions are, in our view, rooted in the distinction between wild and domesticated communities.  We will now review the appraisal of these experiments by the community participants and the managers, which will help to illustrate this.

### 5.1   How the Participants Evaluated the Communities of Practice

At the end of our study, the members of the Project Leaders Community of Practice had participated in seven meetings as a group of individuals, four meetings as a sub-group and had had regular informal exchanges with each other.  They felt that this had greatly improved knowledge sharing and created a feeling of trust.  They realized that they were not the only ones to face the type of problems they had in their day-to-day practice of project management.

The perception of the members of the Technical Experts Community of Practice was less positive.  Ten out of the fifteen regular members felt that nothing had changed.  However, observation of their activity during the first year showed that certain technical problems were solved thanks to the cooperation of experts who had not previously worked together and two new patents could be attributed directly to collaboration between experts.

At the end of the experiment, the members of both communities expressed a wish to continue to work together as a Community of Practice; they also expressed the hope that management would take more account of any future proposals that they made.

## 5.2 How the Company Evaluated the Communities of Practice

The steering committee and the executive committee considered the outcome of the experiment to be positive and the experiment to be satisfactory in providing an answer to the problems of Knowledge Management that had been identified. They recognized that the informal sharing of knowledge could improve practice, but at the same time, also felt that they should be able to measure and evaluate the outcomes; in other words, they wanted to have control over what the communities produced.

The Project Leaders Community of Practice had identified some of the limits of the formal organization. By collectively highlighting problems with existing ways of managing projects, the group had pushed the organization towards a more coherent approach to managing a portfolio of projects. In an echo of Brown and Duguid's tech-reps, they acted *"... to protect the organization from its own shortsightedness"* (Brown & Duguid, 1991, p. 43).

The steering committee was appreciative of the input of the experts' community into the strategic planning process and, in addition, a number of interdisciplinary seminars were organized that contributed to the sharing of knowledge within the organization. The executive committee considered this community to be a valuable resource. However, this recognition also came with a desire to exert more control.

## 6 Discussion and Conclusion

In this article, we have tried to assess, both theoretically and empirically, the implications of the domestication of Communities of Practice. We started by highlighting the shift within the theory related to Communities of Practice: from a view of natural, emergent and creative groups sharing a common practice to a technique for Knowledge Management. We now return to our original question: by cultivating Communities of Practice, do companies risk, in Wenger's terms, killing the goose that lays the golden eggs?

### 6.1 The Benefits of Instrumentalizing Communities of Practice

Our results show that, for both communities, both the participants and the executive committee of the company considered the initiative to have had a positive effect on knowledge sharing. It contributed to the crossing of existing organizational boundaries and an improvement in collective problem solving; it even led to some innovations. Although they faced similar problems, it is unlikely that the either of two Communities of Practice we looked at would have emerged naturally; we believe that a certain degree of instrumentalization or cultivation was necessary.

**Benefits Related To Learning Dynamics**

In our view, outputs such as the new procedures for project management and the mapping of knowledge domains, were effectively boundary objects (Star & Griesemer, 1989) that allowed different groups with diverse interests to work together without needing to establish a formal consensus or specify a set of shared goals. Wenger (1998b) characterized boundary objects as possessing modularity, abstraction, standardization and accommodation. All of these characteristics are displayed in the outcomes described above. They are *modular*, for example, the strategic mapping presented different domains that could be used separately by different technical departments. They are also *abstract* as they abstract away some specific details in order to make the mapping useable by others. In addition, they are *standardized*, for example, project management procedures were described in a standard way so that the different participants in a project would know how to deal with them and finally, they were flexible enough to *accommodate* the practices of various departments such as sales, marketing, engineering, etc.

It appears to us that while the interactions between these communities and the steering committee allowed for the production of a type of knowledge that could be used outside the boundary of the community, the level of control applied by the organization also had the effect of impeding the type of learning associated with emergent or natural Communities of Practice. This created a paradoxical situation where these domesticated Communities of Practice were expected to be both creative and constrained. They found it almost impossible to improvise and produce new ideas or new practices as part of their ongoing stream of activity, because almost everything they wanted to do had to be negotiated with the executive committee. We believe that because these Communities of Practice were formalized and under the constant supervision of the steering committee, they offered no space for what is sometimes termed "*bricolage*" (Jouvenet, 2007).

In summary, we have argued that wild communities produce, more local learning and creativity, contributing to improved local practice through a daily sharing of experience. Domesticated communities, on the other hand, can enhance organizational learning across boundaries through the production of boundary objects but lack the space for improvisation and creativity. To benefit from these communities, we believe that organizations need to allow some autonomy and freedom from routines in order to allow these communities to develop and evolve. This was not the case in the company in our study.

**Benefits Related To Identity Construction**
A positive effect that we observed from the point of view of the participants was related to the construction of a professional identity; this was particularly the case in the expert community. Expert status had only recently been created in the company and the people who were designated as experts also had other functions; it was not clear for them, or for others, what it meant to be an expert.

This is not an example of identity in practice as described by Wenger, where identity arises out of the interplay of participation and reification (Wenger, 1998b, p. 153). Participation in these domesticated communities was low and most of the activities were devoted to reifying existing practices, e.g. through defining sets of common rules. However, Wenger (1998b, p. 173) also describes other modes of belonging: *alignment* (coordinating our energy and activities in order to fit within broader structures), *imagination* (creating images of the world and seeing connections through time and space by extrapolating from our own experience) and *engagement* (active involvement in mutual processes of the negotiation of meaning).

We believe that the community of experts contributed to the creation of a collective identity primarily through the first two types of belonging: alignment and imagination. At first, the experts tried to find some common perspectives. As noted by Wenger (1998b, p. 187), alignment requires the creation and adoption of broader discourse, which is based on reification. It is what the experts did by first trying to better define their status and the content of their duties as experts. In a second phase, when the group was asked to provide a mapping of domains of scientific knowledge, they engaged in a work of imagination. For Wenger, imagination refers to a process of creating new images of the world and ourselves. Consequently, the engagement of the experts in these communal activities created a reality in which they were able to act and construct a shared identity.

This observation suggests that a low level of participation in these communities can be compensated for by other types of belonging, such as alignment and imagination, which can contribute to the construction of an identity. This process is close to that of the Alcoholics Anonymous meetings described by Lave and Wenger (1991). The members do not share the practice of drinking; rather, they try to align their understanding of what it means to be a drinking alcoholic in order to become a non-drinking alcoholic.

### 6.2   The Risks of Instrumentalizing Communities of Practice

Based on the results of the case study, we can put forward three types of risks associated with the instrumentalization of, and overly zealous attempts to control, Communities of Practice.

**Communities That Hide**

We noted previously that it is suggested in the literature that, when a community is over-managed, it may disperse and disappear from view, while still continuing to function 'underground' (Gongla & Rizzuto, 2004). This phenomenon is well-known in the sociology of organizations: people need a space within which they have autonomy and will seek to find ways to protect or extend it (Crozier & Friedberg, 1977). Something similar to this was observed in our expert community. In year 2, it almost ceased to have formal meetings although some of the members continued to pursue 'informal' relationships. Thus, the first risk of cultivating communities is that attempt to hide themselves in order to protect their autonomy.

**Communities That Wither And Die**

We can offer a complementary interpretation of this. As we have suggested, the balance between participation and reification is different between wild and domesticated communities. There is more participation in wild communities, with the risk that not enough knowledge is reified into boundary objects, and there is more reification in domesticated communities, with the risk of killing spontaneity, creativity and the desire to participate. Thus, the second risk of domesticating Communities of Practice is that they will decline and die, not because they face the unwelcome attention of management, but because they are not fed by the participation that nurtures an on-going practice.

**Communities That Go Into Hibernation**

Finally, there is a third possibility. As we noted earlier, there are different types of belonging to the community and different levels of participation. In domesticated Communities of Practice, people need constant encouragement to participate. Thus, in the expert community we observed that when a project, such as the request for a mapping knowledge domains within the company, occurred experts' participation increased, but when their projects were rejected, it fell. Similar observations of communities that form, disappear and then reform have been made elsewhere (Ribeiro, Kimble, & Cairns, 2010). Consequently, the third risk of cultivating domesticating Communities of Practice it is not that the community will die outright, but that it will go into hibernation and need to be revived.

### 6.3   Conclusions and Further Research

From a practical viewpoint, our study has indicated that the domestication of Communities of Practice can bring some benefits, e.g. in terms of learning dynamics and identity construction, but also that it presents some risks, concerning the continued existence of such communities and the problem of how to delegate sufficient autonomy to maintain the motivation of the participants.

We must assume that the benefits observed are contingent to this situation. To some extent, we will have inevitably produced a Hawthorne effect. If you put people together, who did not communicate before and who share an interest, whatever the protocol, you obtain some positive results and improvements in the knowledge sharing process. We might therefore formulate the following hypothesis to explore the implications of out work: "*Do wild communities have more to contribute to an organization than domesticated communities?*"

Wild communities have the potential to produce learning for as long as there is a shared practice, but the challenge is to go beyond the boundaries of the community and link this with the rest of the organization. On the other hand, the domestication of communities could be seen as a first step in bringing people together in order to start some form of exchange of knowledge. We have observed this phenomenon in our case, with some experts and project managers starting to communicate and work with each other outside the community. Thus, perhaps paradoxically, a measure of the success of this type of community would be that it would disappear. At present, we do not know if this collaboration resembles Communities of Practice that are 'in the wild' or are simply a series of bi-lateral relationships.

A second element of context that must be borne in mind is the need for this particular organization to follow a set of extremely rigorous quality procedures. According to Benner and Tushman (2003), quality management procedures are coherent with exploitation but not exploration and innovation. We might expect that in another type of company, perhaps one that is more innovative and less structured, the results of this experiment would have been different. Comparative studies in different types of environment would need to be carried out to ascertain whether the type of company and its culture has any effect on the benefit that can be obtained from the creation of artificial Communities of Practice.

Finally, from the theoretical viewpoint, we believe our work has contributed to a better characterization of the distinction between a wild and a domesticated community. As has been noted by others (Lindkvist, 2005) the term Community of Practice has been stretched to cover a of multitude groups and settings including what would normally be viewed as task groups or teams. We think that it leads to unnecessary ambiguity to use the same concept for both wild and domesticated communities because they are different in nature. In the wild community, participation is at the core; organizational learning only occurs if there is a sufficient level of reification (e.g. through the production of boundary objects) and if the goals of the community are aligned with what might be termed the best interests of the organization. In the domesticated communities, reification is central. The issue then becomes how to obtain and maintain enough participation so that what is reified can be used in the practice.

Valérie Chanal. and Chris Kimble. *Born to be Wild: Using Communities of Practice as a Tool for Knowledge Management*. Paper presented at the Ethicomp 2010: The 'Backwards, Forwards and Sideways' changes of ICT, Tarragona, Spain, April, 2010, pp. 71 - 80.